\documentclass[12pt]{article}
\usepackage{graphicx}
\usepackage{amsmath}
\usepackage{amssymb}
\usepackage{caption2}
\begin{document}
\newcommand  {\ba} {\begin{eqnarray}}
\newcommand  {\be} {\begin{equation}}
\newcommand  {\ea} {\end{eqnarray}}
\newcommand  {\ee} {\end{equation}}
\renewcommand{\thefootnote}{\fnsymbol{footnote}}
\renewcommand{\figurename}{FIG.}
\renewcommand{\captionlabeldelim}{.~}

\begin{flushright}
 {\large\bf BIHEP-TH--2002-31}
\end{flushright}
\vskip 3.0cm

\begin{center}
{\Large\bf Associated production of a Higgs boson with tau sleptons in the \emph{CP} violating MSSM at future
$e^{+}e^{-}$ colliders}
 \vskip 1.0cm
{\bf Wei Min Yang\footnote{Email address: yangwm@mail.ihep.ac.cn} and Dong Sheng Du}\\
 {\em CCAST(World Laboratory), P.O.Box 8730, Beijing 100080, China
 and Institute of High Energy Physics, P.O.Box 918(4), Beijing 100039, China}
 \vskip 2.0cm
\textbf{ABSTRACT}
\end{center}

We investigate the production of the lightest neutral Higgs boson in association with tau sleptons in the
\emph{CP} violating minimal supersymmetric standard model (MSSM) at future high-energy  $e^{+}e^{-}$ linear
colliders. In parameter space of the constrained MSSM, the production cross section of $e^{+}e^{-}\rightarrow h^0
\widetilde{\tau}^+_1\widetilde{\tau}^-_1$ can be very substantial at high energies. This process would provide a
new production mechanism of neutral Higgs bosons and open a window to probe the Higgs-stau
coupling and some soft supersymmetric breaking parameters at next linear colliders (NLC).\\

PACS numbers: 12.60.Jv, 14.80.Cp, 14.80.Ly

\newpage

\begin{center}
\textbf{I. INTRODUCTION}
\end{center}

Supersymmetric (SUSY) theories, in particular the MSSM \cite{1,2}, are currently considered as the most
theoretically well motivated extensions of the standard model (SM). It provides a natural solution to the
hierarchy problem \cite{3}, the generation of the electroweak symmetry breaking \cite{4}, the grand unification of
the gauge coupling constants \cite{5}, as well as the elucidation of the baryon asymmetry and the cold dark matter
in the universe \cite{6a,6b}. The MSSM contains five physical Higgs bosons \cite{7}: two neutral \emph{CP}-even
scalars ($h^0$ and $H^0$), one \emph{CP}-odd neutral scalar $A^0$, and a pair of charged scalars ($H^+$, $H^-$).
Such particles break the electroweak symmetry and are responsible for the masses of the fermions and bosons. The
search for the Higgs bosons and the study of their properties are one of the major goals of present and future
colliders \cite{8}. Contrary to hadron machines, at $e^{+}e^{-}$ colliders the background is relatively clean. A
future high-energy and high-luminosity $e^{+}e^{-}$ linear collider operating at energies $\sqrt{s}=500\sim1000$
GeV would provide an ideal environment for searching the lightest Higgs boson and performing precision
measurements of it \cite{9}.

In the context of the MSSM, some typical production mechanisms of Higgs bosons, such as Higgsstrahlung \cite{10a},
vector bosons fusion \cite{10b}, Higgs pair production \cite{10c}, and associated production of Higgs boson with
top or stop \cite{11a}, have all been extensively studied at NLC energies. In addition, it was realized that
couplings of Higgs bosons to sfermions can also provide additional sources for Higgs bosons production \cite{11b},
in particular, when the mixing between the third generation scalar taus is very large, the associated production
Higgs bosons with tau sleptons might provide a important information on the soft SUSY breaking scalar potential.
In this paper, we will discuss in more detail the last type of the prospecting processes in the \emph{CP}
violating MSSM. Our motivation is as the following: First, for the large Yukawa coupling of the tau slepton, the
mixing between tau sleptons can be very large. As a result, the lighter stau can not only be significantly lighter
than the top quark and all scalar quarks, but also at the same time their couplings to Higgs bosons can be
strongly enhanced. Therefore, the reaction $e^{+}e^{-}\rightarrow h^0 \widetilde{\tau}^+_1\widetilde{\tau}^-_1$
can be more phase-space favored, and also it would be accessible to produce and measure them at the NLC. Second,
it has received growing attention that \emph{CP}-violating phases associated with the third generation trilinear
soft term can induce \emph{CP} violation in the Higgs and the third generation sfermion sectors through quantum
corrections \cite{12}. Such phases may allow baryogenesis and do not necessarily violate the stringent bound from
the nonobservation of electric dipole moments (EDM) \cite{13a,13b}. In fact, some of these phases can be
$\mathcal{O}(1)$, so as to provide non-SM sources of \emph{CP} violation required for dynamical generation of the
baryon asymmetry of the universe \cite{14}. Since these phases directly affect the couplings of Higgs bosons to
third generation sfermions, moreover, these couplings play an important role in the phenomenology of the third
generation sfermions and Higgs bosons at colliders, we will consider these effects of the \emph{CP}-violating
phases on the associated Higgs production with the stau*-stau pair. Last, for most of the SUSY parameter space
allowed by present data constrains, the MSSM Higgs sector is in the decoupling regime \cite{15}. In this scenario,
only the lightest Higgs boson would be accessible at the NLC, final states with the other Higgs bosons or with
$\widetilde{\tau}^+_1\widetilde{\tau}^-_2$ pair should be phase-space suppressed. In what follows, therefore, we
will focus on the case of the lightest $h^0$ boson of MSSM in the decoupling regime, and discuss only the
production in association with lighter tau sleptons.

The remainder of this paper is organized as follows. In Section II we outline the masses and mixing of the tau
sleptons, and then we present the relevant couplings and analytical expressions of the production cross section
for the process $e^{+}e^{-}\rightarrow h^0\widetilde{\tau}^+_1\widetilde{\tau}^-_1$. In Sec.III, a detailed
numerical analysis of the cross section is given in an appropriate MSSM scenario satisfying the experimental and
theoretical constraints. Sec.IV is devoted to the conclusions.

\begin{center}
\textbf{II. MASSES, COUPLINGS AND CROSS SECTION}
\end{center}

In this section, first, to fix notation we will simply summarize the masses and mixing of the tau slepton sector
in the MSSM, and then list the relevant couplings involved in the associated production of the lightest neutral
Higgs boson with tau sleptons. Finally, we will give analytical results of the cross section of
$e^{+}e^{-}\rightarrow h^0\widetilde{\tau}^+_1\widetilde{\tau}^-_1$ production.

\begin{center}
\textbf{A. Tau slepton mass and mixing}
\end{center}

In the tau slepton sector \cite{1,2}, the left- and right-handed tau sleptons can strongly mix, the mass-squared
matrix is given by
 \ba M^2_{\widetilde{\tau}}=\left(\begin{array}{cc}
     M^2_L & m_\tau(A_\tau + \mu\tan\beta) \\
     m_\tau(A_\tau^* + \mu^*\tan\beta) & M^2_R\end{array}\right)
 \ea
 with\\
\parbox{12.9cm}
{\begin{alignat*}{2}
 M^2_L &=m^2_{\widetilde{\tau}_L}+m^2_\tau+(\frac{1}{2}M^2_Z-M^2_W)\cos2\beta\:,\\
 M^2_R &=m^2_{\widetilde{\tau}_R}+m^2_\tau+(M^2_Z-M^2_W)\cos2\beta\:,
\end{alignat*}}
\parbox{0.1cm}{\ba\ea}\\
 where $m_{\widetilde{\tau}_{L,R}}$ are the left- and right-handed soft SUSY-breaking tau slepton masses, respectively.
 The soft breaking trilinear stau coupling $A_\tau$ and Higgs mass mixing parameter $\mu$ are complex,
\ba A_\tau=|A_\tau|e^{i\,\varphi}\:,\hspace{1cm}\mu=|\mu|e^{i\,\eta}\:. \ea
  The phases $\varphi$ and $\eta$ are the sources of \emph{CP} violation. In fact, the phase $\eta$ can be transferred into
   the phase $\varphi$ in the form of $\varphi-\eta$ by the field redefinition. So we will take $\mu$ to be real,
  and vary $\varphi$ from $0$ to $2\pi$ in the following. The tau slepton mass eigenstates can be realized by a
unitary transformation $U$ which diagonalizes the mass-squared matrix $M^2_{\widetilde{\tau}}$\,,
 \ba U\,M^2_{\widetilde{\tau}}\,U^\dagger=\mbox{diag}\,(m^2_{\widetilde{\tau}_1}\,,\,m^2_{\widetilde{\tau}_2})\,,\ea
 where the diagonalization matrix can be parameterized as
 \ba U=\left(\begin{array}{cc}\cos\theta_\tau & \sin\theta_\tau e^{i\,\delta} \\
 -\sin\theta_\tau e^{-i\,\delta} & \cos\theta_\tau\end{array}\right) \ea
 with
 \ba \delta=\arg\,(A_\tau+\mu\tan\beta)\,.\ea
 The tau slepton mixing angle and mass eigenvalues are then given as\\
\parbox{12.9cm}
{\begin{alignat*}{2}
 \tan2\theta &=\frac{2m_\tau|A_\tau+\mu\tan\beta|}{M^2_L-M^2_R}\,,\\
 m^2_{\widetilde{\tau}_{1,2}}
             &=\frac{1}{2}\left[M^2_L+M^2_R\mp\sqrt{(M^2_L-M^2_R)^2+4m^2_\tau|A_\tau+\mu\tan\beta|^2}\right]
\end{alignat*}}
\parbox{0.1cm}{\ba\ea}\\
with convention $0\leqslant\theta\leqslant\pi/2,\,m^2_{\widetilde{\tau}_1}\leqslant m^2_{\widetilde{\tau}_2}$. For
large values of $|A_\tau|$, $|\mu|$ and $\tan\beta$, the mixing in the tau slepton sector can be very strong. In
the constrained MSSM or minimal supergravity (mSUGRA) scenario \cite{1,16}, the soft SUSY breaking scalar fermion
masses and gaugino masses are respectively unified as $m_0$ and $m_{1/2}$ at the grand unified theory (GUT) scale
$M_{GUT}$. The left- and right-handed tau slepton masses at the weak scale ${\cal O}(M_Z)$ are then
given in terms of the universal masses $m_0$ and $m_{1/2}$ through running renormalization group equations (RGE) as \cite{17}\\
\parbox{12.9cm}
{\begin{alignat*}{2}
 m^2_{\widetilde{\tau}_L} & \simeq m_0^2+0.52\, m_{1/2}^2-0.27 M_Z^2 \cos2\beta \,,\\
 m^2_{\widetilde{\tau}_R} & \simeq m_0^2+0.15\, m_{1/2}^2-0.23 M_Z^2 \cos2\beta \,.
\end{alignat*}}
\parbox{0.1cm}{\ba\ea}\\
Inputting values of the parameters $|A_\tau|,\,|\mu|,\,\varphi,\,\tan\beta,\,m_0,\,m_{1/2}$, the mass spectra and
mixing angle of the tau sleptons can then be worked out. For smaller values of $m_0$ and $m_{1/2}$\,, the stau
masses are lower.

\newpage

\begin{center}
\textbf{B. $\mathbf{e^{+}e^{-}\rightarrow h^0\widetilde{\tau}^+_1\widetilde{\tau}^-_1}$ production}
\end{center}

\begin{figure}[t]
\centering
\includegraphics{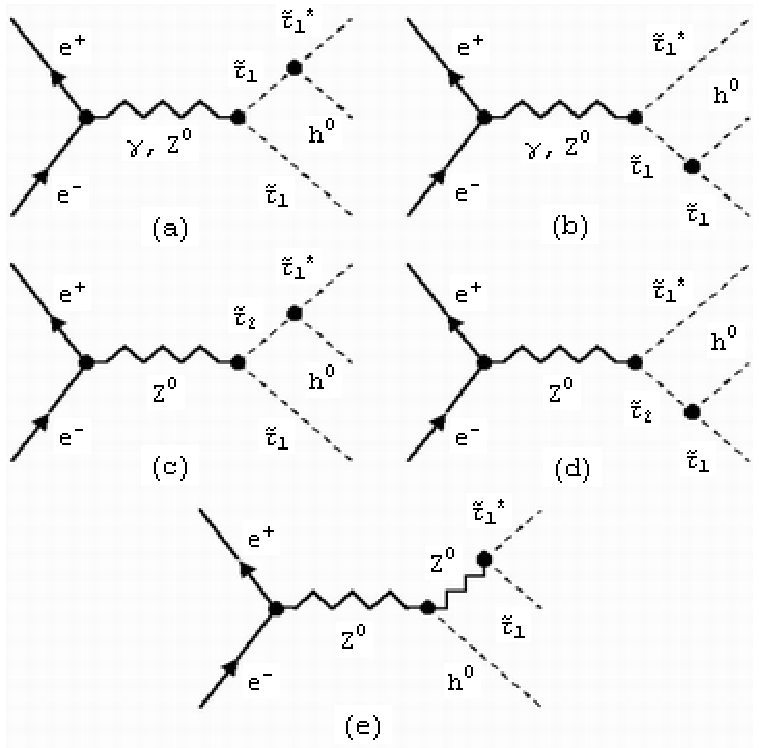}
\caption{Feynman diagrams for $e^{+}e^{-}\rightarrow h^0\widetilde{\tau}^+_1\widetilde{\tau}^-_1$ production.}
\end{figure}
$e^{+}e^{-}\rightarrow h^0\widetilde{\tau}^+_1\widetilde{\tau}^-_1$ production proceeds via three kinds of
diagrams, as shown in Fig.~1. When the $h^0\widetilde{\tau}_1^*\widetilde{\tau}_1$ couplings are large, the most
important is Higgs boson emission from the $\widetilde{\tau_1}$ states which are produced through s-channel photon
and Z-boson exchange (Diag.~a and b in Fig.~1). One also has the conversion of $\widetilde{\tau_2}$ to
$\widetilde{\tau_1}$, which involves the $h^0\widetilde{\tau}_1^*\widetilde{\tau}_2$ vertex (Diag.~c and d in
Fig.~1). The last type is $h^0Z^*$ production with $Z^*\rightarrow\widetilde{\tau}^+_1\widetilde{\tau}^-_1$
(Diag.~e in Fig.~1). Though the two last types of Feynman diagrams turn out to give small contributions, there are
some regions in the parameter space where they cannot be neglected. The relevant couplings for the process are
described by the following set of interaction Lagrangian in the MSSM \cite{1,7}:\\
(a). The couplings of a neutral Higgs boson $h^0$ to a pair of tau sleptons
   $\widetilde{\tau}_i\widetilde{\tau}_j(i,j=1,2)$ is given by
   \ba \mathcal{L}_{h^0\widetilde{\tau}_i^*\widetilde{\tau}_j}=g_{ij}h^0\widetilde{\tau}_i^*\widetilde{\tau}_j \ea
   with
   \begin{alignat}{1}
   g_{ij}=&\frac{g m_Z}{c_W}\left[(s_W^2-\frac{1}{2})\sin(\alpha+\beta)+\left(\frac{m_\tau}{m_Z}\right)^2
          \frac{\sin\alpha}{\cos\beta}\right](U^\dagger_{1i})^* U^\dagger_{1j} \nonumber\\
          &+\frac{g m_Z}{c_W}\left[-s_W^2\sin(\alpha+\beta)+\left(\frac{m_\tau}{m_Z}\right)^2
          \frac{\sin\alpha}{\cos\beta}\right](U^\dagger_{2i})^* U^\dagger_{2j} \nonumber\\
          &+\frac{g m_\tau}{2c_W m_Z}\left[A_\tau^*\frac{\sin\alpha}{\cos\beta}-\mu^*\frac{\cos\alpha}{\cos\beta}\right]
          (U^\dagger_{2i})^* U^\dagger_{1j} \nonumber\\
          &+\frac{g m_\tau}{2c_W m_Z}\left[A_\tau\frac{\sin\alpha}{\cos\beta}-\mu\frac{\cos\alpha}{\cos\beta}\right]
          (U^\dagger_{1i})^* U^\dagger_{2j} \,,
   \end{alignat}
   where $s_W=\sin\theta_W,c_W=\cos\theta_W$, $\alpha$ is the \emph{CP}-even Higgs mixing angle at the tree level.
   When the loop corrections are included, the tree level parameters are deformed, even inducing the mixing among
   Higgs with different \emph{CP} properties \cite{18}. In the decoupling limit and up to very small radiative corrections,
   however, the mixing angle $\alpha$ reaches the limit $\beta-\pi/2$. \\
(b). The couplings of neutral vector gauge bosons ($\gamma$ and $Z^0$) to tau slepton pairs are written as
   \ba \mathcal{L}_{V\widetilde{\tau}_i^*\widetilde{\tau}_j}=\frac{ig}{c_W}\left[s_W c_W\delta_{ij}A^\mu
      +\left(-s_W^2\delta_{ij}+\frac{1}{2}(U^\dagger_{1i})^* U^\dagger_{1j}\right)Z^\mu\right]
       \widetilde{\tau}_i^* \stackrel{\leftrightarrow}{\partial}_\mu \widetilde{\tau}_j \,. \ea
(c). The couplings of neutral vector gauge bosons to charged lepton pairs are described by the same one as in the
     SM
   \ba \mathcal{L}_{V\overline{f}f}=\frac{g}{c_W}\left\{s_W c_W A^\mu \overline{f}\gamma_{\mu}f
       +Z^\mu\overline{f}\gamma_{\mu}\left[\left(\frac{1}{2}-s_W^2\right)\frac{1-\gamma^5}{2}
       -s_W^2\frac{1+\gamma^5}{2}\right]f\right\}. \ea
(d). The couplings of a neutral Higgs boson to a pair of massive neutral gauge bosons is given by
   \ba \mathcal{L}_{h^0ZZ}=\frac{g m_Z}{2c_W}\sin(\beta-\alpha) h^0 Z^{\mu}Z_{\mu} \,. \ea
Using Feynman rules, the Dalitz plot density of the production cross section can be obtained. In terms of the
energies $E_1$, $E_2$, $E_h$ of the three final state particles $\widetilde{\tau}_1^-$, $\widetilde{\tau}_1^+$ and
$h^0$, respectively, it is written as
\begin{alignat}{1}
\frac{\mbox{d}\sigma}{\mbox{d}E_1\mbox{d}E_2}= &\frac{m_e^2}{8(2\pi)^3\sqrt{s^2-4sm_e^2}}\left\{\frac{s}{8}|A|^2
+\frac{\sqrt{s}E_h}{4}\mbox{Re}(AB^*)\right. \nonumber\\
&\left.+\left[\frac{m_h^2}{8}+\left(\frac{1}{8}-\frac{1}{2}s_W^2+s_W^4\right)\left(E_h^2-m_h^2\right)\left(\frac{s}{3m_e^2}+\frac{2}{3}\right)\right]|B|^2\right\}\,,
\end{alignat}
where as usual $s$ is the total center-of-mass energy squared, the coefficients $A$ and $B$ are defined as
follows:
\begin{alignat}{1}
A=&\left(\frac{g}{c_W m_Z}\right)^2\left[g_{11}\left(\frac{|U^\dagger_{11}|^2}{2}-s_W^2\right)
   \left(\frac{1}{s-2\sqrt{s}E_1}+\frac{1}{s-2\sqrt{s}E_2}\right)\right. \nonumber\\
  &\left.+g_{12}\frac{(U^\dagger_{12})^*U^\dagger_{11}}{2}\frac{1}{s-2\sqrt{s}E_2+m_1^2-m_2^2}\right. \nonumber\\
  &\left.+g_{21}\frac{(U^\dagger_{11})^*U^\dagger_{12}}{2}\frac{1}{s-2\sqrt{s}E_1+m_1^2-m_2^2}\right. \nonumber\\
  &\left.+i\left(\frac{|U^\dagger_{11}|^2}{2}-s_W^2\right)\frac{g\sin(\beta-\alpha)(s-\sqrt{s}E_h-m_Z^2)}
   {c_W m_Z(s-m_Z^2)}\right]\,, \nonumber\\
B=&\,i\left(\frac{|U^\dagger_{11}|^2}{2}-s_W^2\right)\frac{g^3\sin(\beta-\alpha)}{c_W^3 m_Z(s-m_Z^2)}\,.
\end{alignat}
To obtain the total production cross section $\sigma(e^{+}e^{-}\rightarrow
h^0\widetilde{\tau}^+_1\widetilde{\tau}^-_1)$, one can rewrite equation (14) in the standard form for the Dalitz
plot density \cite{19}, and then integrate over the variables $s_{12}$, $s_{23}$ whose integration bounds are
given from kinematics. For a given center of mass energy the cross section depends on the masses of created
particles and vanishes at the kinematic boundary.

\begin{center}
\textbf{III. NUMERICAL RESULTS}
\end{center}

In this section, we will illustrate our numerical results of the total production cross section for
$e^{+}e^{-}\rightarrow h^0\widetilde{\tau}^+_1\widetilde{\tau}^-_1$ based on a specific constrained MSSM scenario
for the relevant SUSY parameters. The whole parameters appearing in our analyses contain
$|A_\tau|,\,|\mu|,\,\varphi,\,\tan\beta,\,m_0,\,m_{1/2},\,m_h$. Including the two-loop radiative corrections, the
mass of the lightest Higgs boson, $m_h$, depends mainly on the following variables \cite{20}: the top quark mass,
the top and bottom squark masses, the mixing in the stop sector, the pseudoscalar Higgs mass $m_A$ and
$\tan\beta$. In particular scenarios, like in SUGRA or gauge and anomaly mediation \cite{16,21a,21b}, some of the
soft parameters may be related to each other. However, since the mechanism of SUSY breaking is unknown, the soft
SUSY breaking parameters associated with the tau slepton sector can be considered to be independent of the
parameters in the Higgs and squark sector \cite{22}. In what follows, therefore, we consider the former seven
parameters as free phenomenological parameters to be constrained by experimental and theoretical considerations.
Among the constraints that we are going to impose on the MSSM are those which follow from the comparison of the SM
with experimental data, from the experimental limits on the masses of yet unobserved particles \cite{19}, etc, and
also those that follow from the ideas of unification and from SUSY GUT models \cite{23}.

\begin{figure}
\centering
\includegraphics[totalheight=5.5cm]{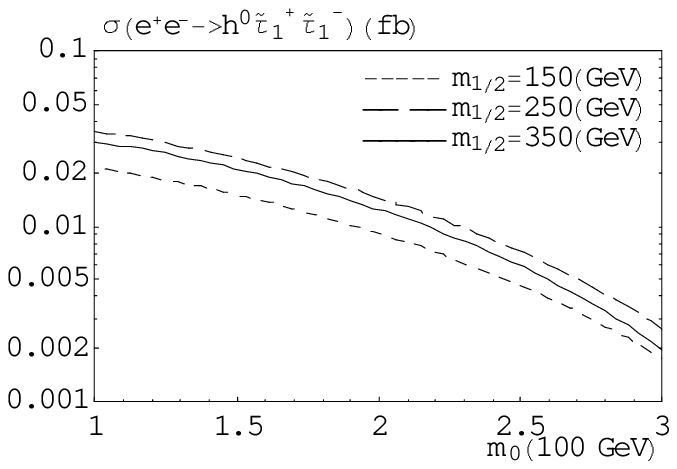}
\caption{The cross section $\sigma(e^{+}e^{-}\rightarrow h^0\widetilde{\tau}^+_1\widetilde{\tau}^-_1)$ as a
function of the scalar fermion mass $m_0$ for $\tan\beta=5$, $|\mu|=500$ GeV, $|A_\tau|=1$ TeV, $\varphi=\pi/4$,
$m_h=120$ GeV. Three curves correspond respectively to three representative values of the gaugino mass
$m_{1/2}$\,.}
\end{figure}
In the following numerical analysis, we fix the center of mass energy to $\sqrt{s}=1$ TeV, and then demonstrate in
turn the dependence of the cross section on the choice of parameters. In Fig.~2 and Fig.~3, the cross section is
shown as a function of the unified scalar fermion mass $m_0$ for the case of low and high $\tan\beta$ value:
$\tan\beta=5$ and $\tan\beta=40$, respectively. Three curves in each figure correspond respectively to three
chosen representative values of the unified gaugino mass $m_{1/2}$\,. The other parameters are typically taken,
for example, as the following: $|A_\tau|=1$ TeV, $\varphi=\pi/4$, $m_h=120$ GeV, $|\mu|=500$ GeV (for
$\tan\beta=5$), $|\mu|=300$ GeV (for $\tan\beta=40$). As can be seen, a typical magnitude of the cross section is
$0.002\sim0.03$ fb. The variation of magnitude depends mainly on $m_0$\,. As the value of $m_0$ is increased, the
cross section drops rather dramatically. The values of $m_{1/2}$ do not change the whole trends of the plots, but
can slightly shift the value of the cross section in the same order of magnitude. This is due to the fact that the
tau slepton masses basically depend on $m_0$\,. In addition, the value of $\tan\beta$ can also influence the value
of the cross section obviously, which can be seen from the position interchange of the solid line and the
long-dashed line in the two graphs, in particular, for the large $\tan\beta$, smaller $m_0$ and $m_{1/2}$ are
excluded by experimental limits. As the value of $\tan\beta$ increases from low to large, the cross section
decreases significantly. However, for no more than about 200 GeV value of $m_0$ in the case of the low $\tan\beta$
(about 150 GeV in the case of the large $\tan\beta$), the production cross section can exceed the value
$\sigma(e^{+}e^{-}\rightarrow h^0\widetilde{\tau}^+_1\widetilde{\tau}^-_1)\sim0.01$ fb. With a integrated
luminosity 500 $\mbox{fb}^{-1}$, this provides several dozen events in a few years, which should be sufficient as
a sample to isolate the final state and measure the $g_{h^0\widetilde{\tau}_1^*\widetilde{\tau}_1}$ coupling with
some accuracy.
\begin{figure}
\centering
\includegraphics[totalheight=5.5cm]{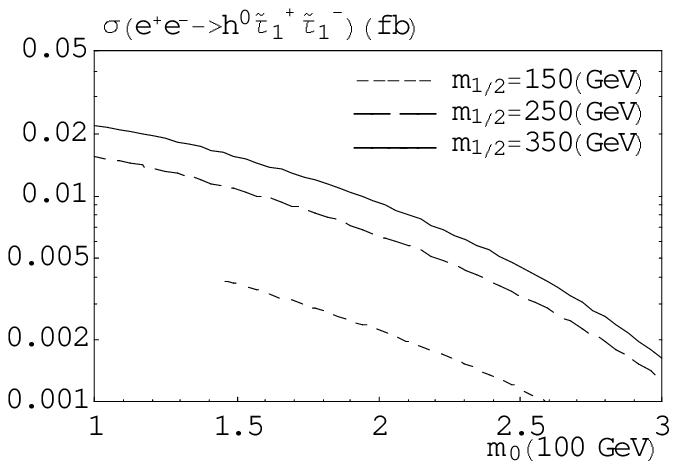}
\caption{The cross section $\sigma(e^{+}e^{-}\rightarrow h^0\widetilde{\tau}^+_1\widetilde{\tau}^-_1)$ as a
function of the scalar fermion mass $m_0$ for $\tan\beta=40$, $|\mu|=300$ GeV, $|A_\tau|=1$ TeV, $\varphi=\pi/4$,
$m_h=120$ GeV. Three curves correspond respectively to three representative values of the gaugino mass
$m_{1/2}$\,.}
\end{figure}

In the Fig.~4, the cross section is displayed as a function of the Higgs mass mixing parameter $|\mu|$ for three
representative values of the lightest Higgs boson mass $m_h$\,, and the other parameters are fixed to
$\tan\beta=5$, $|A_\tau|=1$ TeV, $\varphi=\pi/4$, $m_0=200$ GeV, $m_{1/2}=250$ GeV. The value of $|\mu|$ is
constrained by the requirement of radiative electroweak symmetry breaking \cite{4}, while the values of $m_h$ are
chosen from the current experimental and theoretical considerations \cite{20,24}. The plots show that the cross
section changes somewhat large with values of $|\mu|$ but is not too sensitive to the values of $m_h$. For a
larger value of $|\mu|$, the cross section is smaller. When the value of $m_h$ is increased by 20 GeV, the cross
section drops approximately 0.001 fb.
\begin{figure}
\centering
\includegraphics[totalheight=5.5cm]{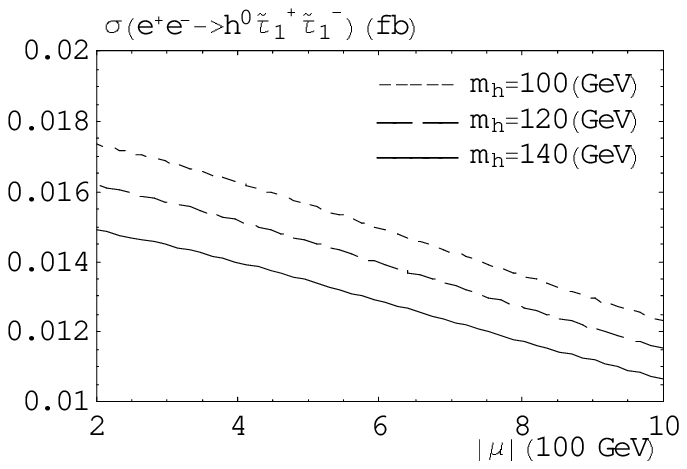}
\caption{The cross section $\sigma(e^{+}e^{-}\rightarrow h^0\widetilde{\tau}^+_1\widetilde{\tau}^-_1)$ as a
function of the Higgs mass parameter $|\mu|$ for $\tan\beta=5$, $|A_\tau|=1$ TeV, $\varphi=\pi/4$, $m_0=200$ GeV,
$m_{1/2}=250$ GeV. Three curves correspond respectively to three representative values of the lightest Higgs boson
mass $m_h$\,.}
\end{figure}

\begin{figure}
\centering
\includegraphics[totalheight=5.5cm]{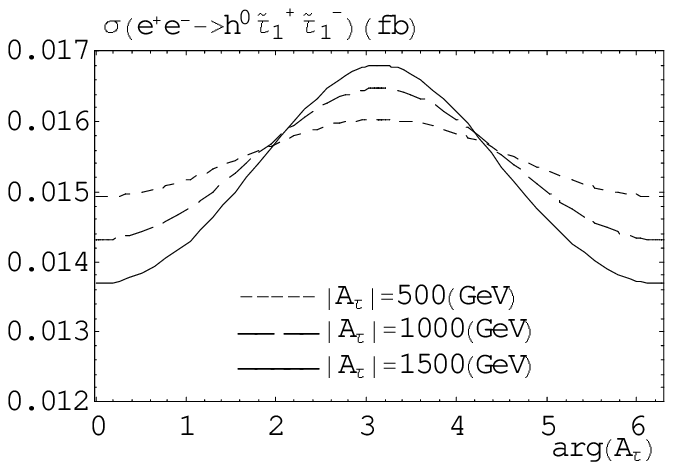}
\caption{The cross section $\sigma(e^{+}e^{-}\rightarrow h^0\widetilde{\tau}^+_1\widetilde{\tau}^-_1)$ as a
function of the \emph{CP}-violating phase $\varphi$ for $\tan\beta=5$, $|\mu|=500$ GeV, $m_h=120$ GeV, $m_0=200$
GeV, $m_{1/2}=250$ GeV. Three curves correspond respectively to three representative values of the soft breaking
trilinear stau coupling $|A_\tau|$\,.}
\end{figure}
The \emph{CP} violating phases in $A_{\tau}$, arg($A_{\tau}$), can modify the masses and the mixing parameters of
stau as well as the Higgs-stau*-stau couplings. This would also have large impact on the production cross section
of the $h^0\widetilde{\tau}^+_1\widetilde{\tau}^-_1$ final state since it is directly proportional to the square
of the $h^0\widetilde{\tau}^+_1\widetilde{\tau}^-_1$ couplings. In Fig.~5, the cross section is shown as a
function of the \emph{CP}-violating phase $\varphi$ (=arg($A_\tau$)) for three representative values of the soft
breaking trilinear stau coupling $|A_\tau|$\,, and the other parameters are fixed to $\tan\beta=5$, $|\mu|=500$
GeV, $m_h=120$ GeV, $m_0=200$ GeV, $m_{1/2}=250$ GeV. The graph shows that the cross section has a symmetry about
$\varphi=\pi$. This results from that the stau mass $m_{\widetilde{\tau}_1}$ are symmetric under
$\varphi\rightarrow\varphi-2\pi$. Furthermore, the cross section reaches a maximum near $\varphi=\pi$. This is
because the coupling $g_{h^0\widetilde{\tau}_1^*\widetilde{\tau}_1}$ has a maximum at $\varphi=\pi$ and its phase
dependence is stronger than that of the masses of stau. However, the trends of plots and the varying magnitude of
the cross section are obviously subject to the values of $|A_\tau|$, in particular, it should be noted that the
cross section might increase or also decrease with the values of $|A_\tau|$ in deferent interval of the phase
$\varphi$.

\begin{center}
\textbf{IV. CONCLUSIONS}
\end{center}

In summary, in the framework of the \emph{CP} violating MSSM, we have discussed the production of the lightest
neutral Higgs particle in association with the lighter tau slepton pair, $e^{+}e^{-}\rightarrow
h^0\widetilde{\tau}^+_1\widetilde{\tau}^-_1$, at future high-energy $e^{+}e^{-}$ linear colliders. Simple
analytical formulae for the cross section of the three-body process have been given. In a specific constrained
MSSM scenario, where the lightest Higgs boson is in the decoupling regime, we have analyzed in detail the
dependence of the cross section on the relevant SUSY parameters. On the one hand, for large values of $|A_\tau|$,
$|\mu|$ and $\tan\beta$, the mixing between tau sleptons can be very large, this results in the lighter stau can
not only be rather light but also at the same time their couplings to Higgs bosons can be strongly enhanced. As a
result, the cross section can be very substantial. On the other hand, the effect of the \emph{CP} violating phase
in $A_\tau$ on the cross section is also demonstrated. It can induce a sizable change of the cross section through
the significant dependence of the stau masses and maxing as well as the
$h^0\widetilde{\tau}^+_1\widetilde{\tau}^-_1$ couplings on the phase. At $e^{+}e^{-}$ colliders with c.m.energy
around $\sqrt{s}=1$ TeV and with very high luminosity $\int\mathcal{L}\mbox{d}t\sim500$ $\mbox{fb}^{-1}$, the
typical cross section is of the order of $\mathcal{O}(0.002\sim0.03)$ fb depending mainly on the universal scalar
fermion mass $m_0$, in particular for light $m_0$ ($m_0\lesssim150$ GeV), the cross section can exceed the level
of a 0.01 fb, leading to several dozen events in a few years. In the case where the tau slepton decays into a
neutrino and a chargino, the final state topology should be easy to be seen experimentally owing to the clean
environment of these colliders. Therefore, all of these results can be expected to be detectable at future
$e^{+}e^{-}$ linear colliders. Analyzing these final states will allow to measure this important coupling and
determine the value of the \emph{CP} violating phase, thus opening a window to probe directly some of the
soft-SUSY breaking scalar potential and also to shed some light on the SUSY breaking mechanism.

\begin{center}
\textbf{ACKNOWLEDGMENTS}
\end{center}

One of the authors, W. M. Yang, thanks A. Djouadi for useful comments very much, and thanks M. Z. Yang for helpful
discussions. This work is in part supported by National Natural Science Foundation of China.

\end{document}